\documentclass{jaa}

\usepackage{graphicx}

\begin{document}

\title{Observational aspects of Outbursting Black Hole Sources - Evolution of
Spectro-Temporal features and X-ray Variability}


\author{Sreehari H.\textsuperscript{1,2,*}, Anuj Nandi\textsuperscript{1}, Radhika D.\textsuperscript{3}, Nirmal Iyer\textsuperscript{4} \and Samir Mandal\textsuperscript{5}}
\affilOne{\textsuperscript{1}Space Astronomy Group, ISITE campus, ISRO Satellite Centre, Outer Ring Road, Marathahalli, Bengaluru, 560037, India.\\}
\affilTwo{\textsuperscript{2}Department of Physics, Indian Institute of Science, Bengaluru, 560012, India.\\}
\affilThree{\textsuperscript{3}Department of Physics, Dayananda Sagar University, Bengaluru, 560068, India.\\}
\affilFour{\textsuperscript{4}Albanova university centre, KTH PAP, Stockholm 10691, Sweden.\\}
\affilFive{\textsuperscript{5}Department of Physics, Indian Institute of Space Science and Technology, Trivandrum, 695547, India.\\}


\twocolumn[{

\maketitle

\corres{hjsreehari@gmail.com}

\msinfo{28 August 2017}{28 August 2017}{28 August 2017}

\begin{abstract}
We report on our attempt to understand the outbursting profile of Galactic Black Hole (GBH) 
sources, keeping in mind the evolution of temporal and spectral features during the outburst.
We present results of evolution of Quasi-periodic Oscillations (QPOs), spectral states
and possible connection with Jet ejections during the outburst phase. Further, we attempt to
connect the observed X-ray variabilities (i.e., `class' / `structured' variabilities, similar
to GRS 1915+105) with spectral states of BH sources. Towards these studies, we
consider three Black Hole sources that have undergone single (XTE J1859+226), a few
(IGR J17091-3624) and many (GX 339-4) outbursts since the start of {\it RXTE} era. Finally, 
we model the broadband energy spectra ($3 - 150$ keV) of different spectral states using
{\it RXTE} and {\it NuSTAR} observations. Results are discussed in the context of two
component advective flow model, while constraining the mass of the three BH sources.

\end{abstract}

}]


\doinum{12.3456/s78910-011-012-3}
\artcitid{\#\#\#\#}
\volnum{123}
\year{2017}
\pgrange{1--12}
\setcounter{page}{1}
\lp{12}


\section{Introduction}

A Galactic Black Hole (GBH) binary system consists of a primary black hole and a secondary star. 
If the secondary star is a Sun like star with mass of a few solar masses, it is called a Low 
Mass X-ray Binary (LMXB) and if the companion star is a few tens of solar masses in size, it is 
a High Mass X-ray Binary (HMXB). The primary accretes matter from the secondary star via
Roche lobe over flow in the case of LMXBs and from stellar winds in the case of HMXBs. A few 
X-ray binaries also have intermediate mass companions and such systems are termed 
Intermediate Mass X-Ray binaries (IMXBs) (Podsiadlowski et al. 2002). 

Tetarenko {\em et al.} (2016) has reported 77 GBH sources while Corral-Santana {\em et al.} (2015) 
reported 59 transient GBH sources of which 18 are dynamically confirmed. Till date, three 
extragalactic BH transients also have been observed. They are LMC X-1, LMC X-3 and M33 X-7 
(see Corral-Santana {\em et al.} (2015) for details). Besides these, there are Ultra-luminous 
X-ray sources (ULXs) which may harbour intermediate mass black holes (Feng \& Soria 2011). 
GBH systems can be either persistent or outbursting (Tanaka \& Shibazaki 1996). Some GBHs like 
Cyg X-1 shows persistent X-ray emission while some others like GRS 1915+105 are persistent with 
aperiodic X-ray variability. Sources like GX 339-4 undergo frequent outbursts separated by 
quiescent phases. There are also sources like GRO J1655-40 that remain mostly in the quiescent 
phase and goes into outburst once in a decade or so. 

During an outburst a typical black hole binary system goes through different canonical states 
(Homan \& Belloni 2005; Belloni {\em et al.} 2005; Remillard \& McClintock 2006; 
Nandi {\em et al.} 2012). In general, states are classified as Low-Hard State (LHS), 
Hard-Intermediate State (HIMS), Soft-Intermediate State (SIMS) and the High-Soft State (HSS). 
It is the mass transfer rate onto the BH which determines the transient behaviour of the source 
(Tanaka \& Lewin 1995). During the LHS, the source energy spectrum can be modelled mainly with 
a \textit{powerlaw} component along with a \textit{gaussian} (for the iron line) and reflection 
components. The presence of a disk is usually seen as the system undergoes transition into the 
intermediate states and the disk contribution increases significantly as the source enters the 
HSS. Meanwhile, the \textit{powerlaw} index also increases from around 1.5 in the LHS to values 
as high as 2.8 in the HSS.

One can study the state evolution of BH binaries associated with the various branches of the 
`q-diagram' or Hardness Intensity Diagram (HID) observed during outburst. The HID of a typical 
outburst is a hysteresis loop in the shape of a `q' (Homan \& Belloni 2005). From the HID, 
it is evident that the decay phase of the outburst always has lower total flux values as 
compared to the rising phase.

As the system evolves through different states, variation in its temporal characteristics has been
observed. Specifically, we see different types of Quasi-periodic Oscillations (QPOs) in the power 
spectrum of the source. Low Frequency QPOs (LFQPOs) are generally found in the LHS, HIMS and SIMS. 
They are categorised into Type A, Type B and Type C based on the frequency of oscillation, 
rms value, quality factor, and significance. Casella {\em et al.} (2004) has given a detailed 
classification of LFQPOs. A few black hole binary systems also exhibit High Frequency QPOs 
(HFQPOs). But in this paper, we restrict our study only to C-Type LFQPOs as they show evolution 
in their frequencies during the rising and decay phases of an outburst.

BH binaries (BHBs) also exhibit different types of X-ray variability in their light curves. Some 
sources like GRS 1915+105 are known to exhibit `structured' or `class' variability. Apart from 
X-ray outbursts and variabilities in smaller time scales, BHBs are also observed to exhibit 
bipolar radio jet emissions (Fender {\em et al.} 2004). 
 
\begin{figure*}
\includegraphics[height=6cm, width=\textwidth]{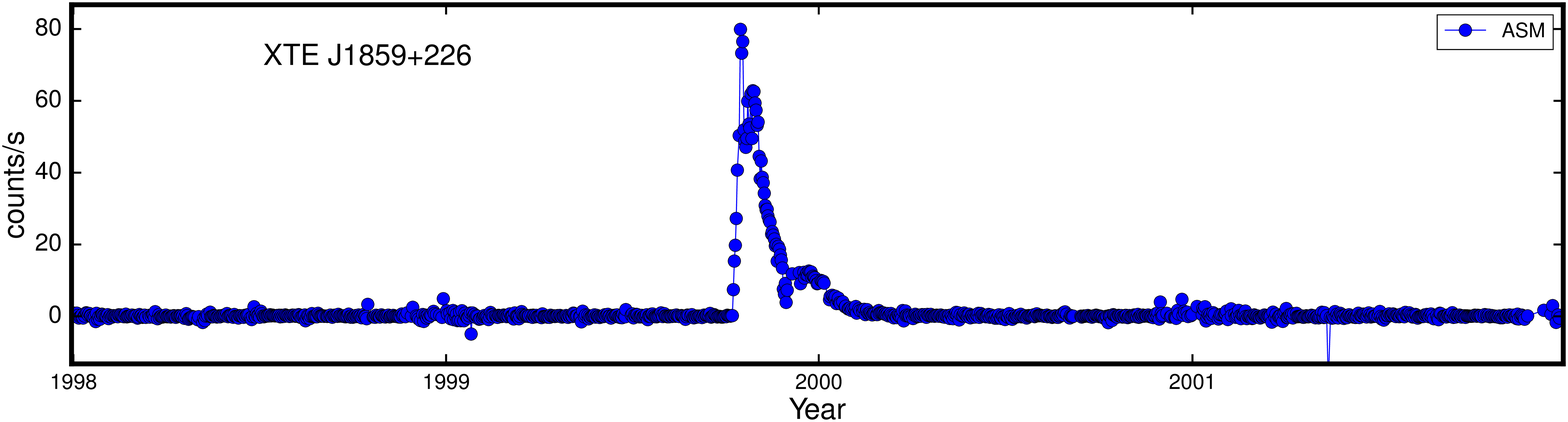}
\includegraphics[height=6cm, width=\textwidth]{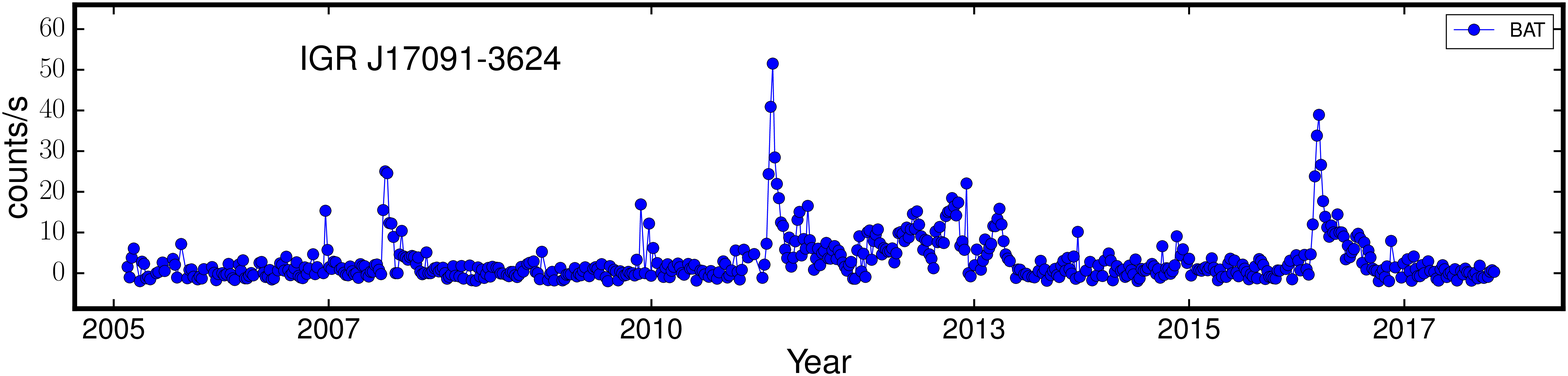}
\includegraphics[height=6cm, width=\textwidth]{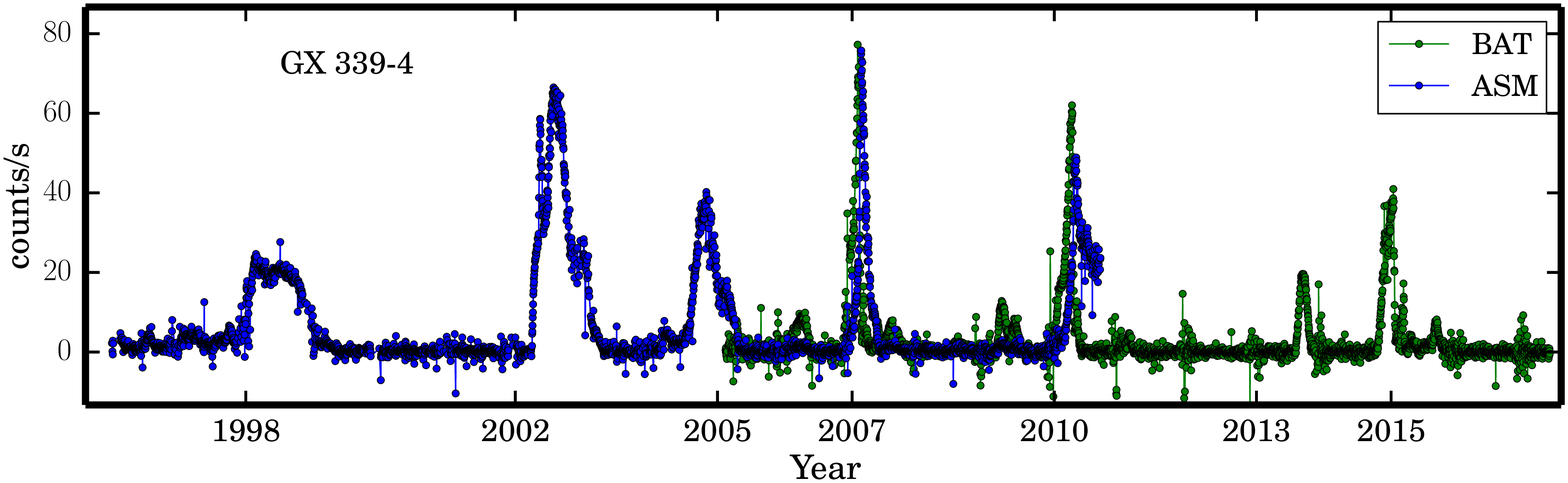}
\caption{Light curve of some outbursting GBHs showing that the frequency of outburst varies from 
source to source. XTE J1859+226 has had only one outburst in the last two decades. GX 339-4 goes 
into outburst once in every two or three years while IGR J17091-3624 clearly shows outbursts in 
2007, 2011 and 2016.} 
\label{fig:outbursts}
\end{figure*}

\subsection{Motivation and Source Selection}

The motivation of this work is to do a comparative study of the outburst profiles, QPO evolution 
and evolution of X-ray spectral features of three BH binary sources namely XTE J1859+226, 
GX 339-4 and IGR J17091-3624. These three sources have also shown signature of X-ray variability 
during their outburst phases. We compare these variability with the structured variability 
exhibited by the source GRS 1915+105.

XTE J1859+226 has gone into an outburst only in 1999 following which it has been in quiescence. 
GX 339-4 is one of the most active transient sources which has undergone several outbursts in 
the {\it RXTE-SWIFT} era. Meanwhile, IGR J17091-3624 undergoes an outburst approximately once in 
every four years since its discovery in 2003. Hence, we have chosen these three BHBs which differ 
in their frequency of outbursts. Figure \ref{fig:outbursts} shows the frequency of outbursts 
of three different BH sources. 

Besides this all three sources have shown evolution of Low Frequency QPOs and we could model 
them with the Propagating Oscillatory Shock (POS) model. Also, we have done spectral modelling 
of the four different states of the BH binary GX 339-4 using two component advective flow model. 
For this, we have chosen only the 2002 outburst of GX 339-4 as a representative case. The 
modelling with phenomenological models as well as with the two component flow for the hard and 
soft states of the other two sources (XTE J1859+226 and IGR J17091-3624) are also shown in 
this paper. Estimation of mass of the black holes for these sources are also done using the 
two component flow model. Detailed modelling of the entire outbursts of these two sources is 
going on and will be published elsewhere. We have also compared the variability exhibited by 
the three sources we consider with that exhibited by GRS 1915+105, which is known for the 
structured variability that it exhibits.  

We also intend to study the possible connection between X-ray variability and state classification 
of the sources there by interpreting the state of GRS 1915+105. This is done based on the states 
of the three BHBs (XTE J1859+226, GX 339-4 and IGR J17091-3624) during the period in which 
they exhibit variability.

\section{Observation, Analysis and Modelling}

We have considered the observations of different BH binaries carried out with the All Sky Monitor 
(ASM), Proportional Counter Array (PCA) and the High Energy X-ray Timing Experiment (HEXTE) 
aboard the satellite Rossi X-ray Timing Explorer (RXTE). The ASM operated in the energy range 
of 1.3 keV to 12.2 keV and we obtain the outburst profile of various sources using data from
this instrument. PCA data in the energy range of 2 to 60 keV and HEXTE data in 15 to 200 keV were 
usually considered for spectral studies. We extract the energy spectrum from PCA and HEXTE using 
standard techniques. The PCA data is used for both spectral and temporal analysis while HEXTE 
is used mainly for generating energy spectrum at higher energies. Broadband energy spectrum 
combining PCA and HEXTE is modelled to study the state evolution of different BH binary systems. 

Besides the spectral studies, we also search PCA data for the presence of peaked noise components 
or QPOs. The science event files are used to extract the light curves which are used to generate 
the power spectrum. We model the obtained power spectra with \textit{powerlaws} and/or 
\textit{Lorentzians}. If any QPO like feature is found then it is fit with \textit{Lorentzians}. 
The significance of QPOs in the spectrum are calculated as the ratio of norm to its negative 
error. If significance $>$ 3 and the Quality factor, $Q=\nu/FWHM$ is found to be greater than 2 
then the narrow feature is considered as a QPO. QPOs are classified into low frequency QPOs 
(LFQPOs) and high frequency QPOs (HFQPOs) based on their frequency ranges. Typically LFQPOs 
appear in the frequency range of 0.1 to 30 Hz and HFQPOs appear at frequencies more than 30 Hz 
and extends upto $\approx$ 500 Hz in the case of BH binaries. The highest HFQPO observed so far 
in BH binaries is from the source GRO J1655-40 and it is $\sim$ 450 Hz (Strohmayer 2001).

We also analysed \textit{SWIFT - XRT} data of IGR J17091-3624 and studied the variability of 
the source. \textit{XRT} operates in the range from 0.5 to 10 keV. \textit{xrtpipeline} was 
run to obtain cleaned event files which were then filtered corresponding to grades 0-2 
using \textit{XSELECT} software. We chose a circular region of 30 arc seconds centred at the 
source RA and DEC to obtain the source region and an annular region of 60 arc second inner 
radius and 90 arc second outer radius as the background region (see Radhika {\em et al.} 2016b 
for further details). Then we extract light curves and energy spectra corresponding to these 
source and background regions. The background subtracted light curves are then used to produce 
the power spectra for further analysis. We use \textit{xrtmkarf} to generate the \textit{arf} 
files and then we re-bin the extracted source energy spectra to 25 counts per bin with the 
ftool \textit{grppha}. 

Six TOO observations with NuSTAR are also available for the 2016 outburst of IGR J17091-3624. 
We include spectral studies based on two of those (from LHS and SIMS) in this paper. The other 
four broadband spectra are studied in detail by Radhika et al. (2018). We use \textit{nupipeline} 
to extract the level 2 data based on the procedure given in NuSTAR guide 
\footnote{https://heasarc.gsfc.nasa.gov/docs/nustar/analysis/nustar swguide.pdf}. A circular 
region of $30^{\prime \prime}$ centred at the source RA and DEC is used as the source region 
and another $30^{\prime \prime}$ circle far away from the source is taken as the background 
region. Then we use the ftool \textit{nuproducts} to extract the spectrum, response and arf 
files. The obtained spectrum is re-binned to 30 counts per bin. In the next section, we present 
the results of the data analysis and modelling done with the three sources under consideration.

\section{Results: Spectro-Temporal features of Outbursting BH Sources}

In order to study the spectro-temporal features of outbursting BH sources, we consider three BH 
sources, namely XTE J1859+226, GX 339-4 and IGR J17091-3624. The X-ray transient source 
XTE J1859+226 was discovered with ASM onboard RXTE on Oct 9, 1999 (Wood {\em et al.} 1999). 
Detailed studies of the outburst had revealed the spectral state characteristics of the source 
(Homan \& Belloni 2005; Radhika \& Nandi 2014). GX 339-4 is a well known transient source and 
has undergone several outbursts during the $RXTE-SWIFT$ era (See Figure \ref{fig:outbursts}). 
Detailed study of its spectral and temporal evolution have paved way for understanding the 
general characteristics of GBH transients (Motta {\em et al.} 2011; Nandi {\em et al.} 2012). 
The source IGR J17091$-$3624 was discovered by the International Gamma-ray Astrophysics 
Laboratory (INTEGRAL) in the year 2003 and also exhibited multiple outbursts. A very detailed 
study of the source has revealed its spectral and temporal properties 
(Capitanio {\em et al.} 2006; Capitanio {\em et al.} 2012; Iyer {\em et al.} 2015). The most 
significant feature of the source is the variabilities/oscillations discovered in its light 
curve during the 2011 outburst (Altamirano {\em et al.} 2011). These have been found to be 
similar to the well known source GRS 1915+105 (Belloni {\em et al.} 2001). In the following 
subsections, we present results and discuss the Outburst profile, evolution of QPO frequencies, 
Spectral state evolution, Jet connection with spectral states, X-ray variability and broadband 
spectral modelling based on two component flow model of BH binary sources. 

\begin{figure}[!hbtp]
\includegraphics[height=8cm,width=\columnwidth]{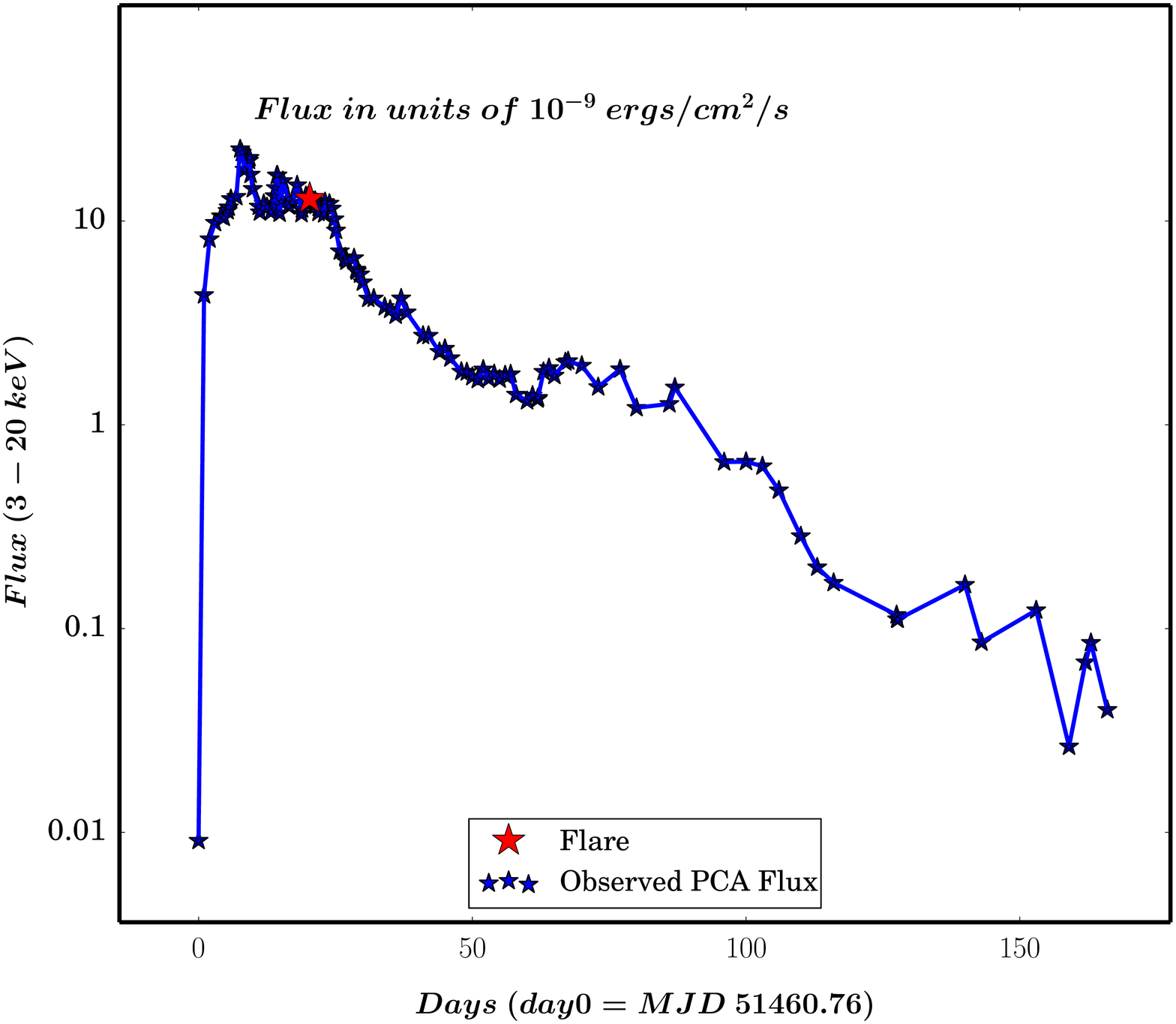}
\begin{flushright}
\includegraphics[height=8cm,width=0.9\columnwidth]{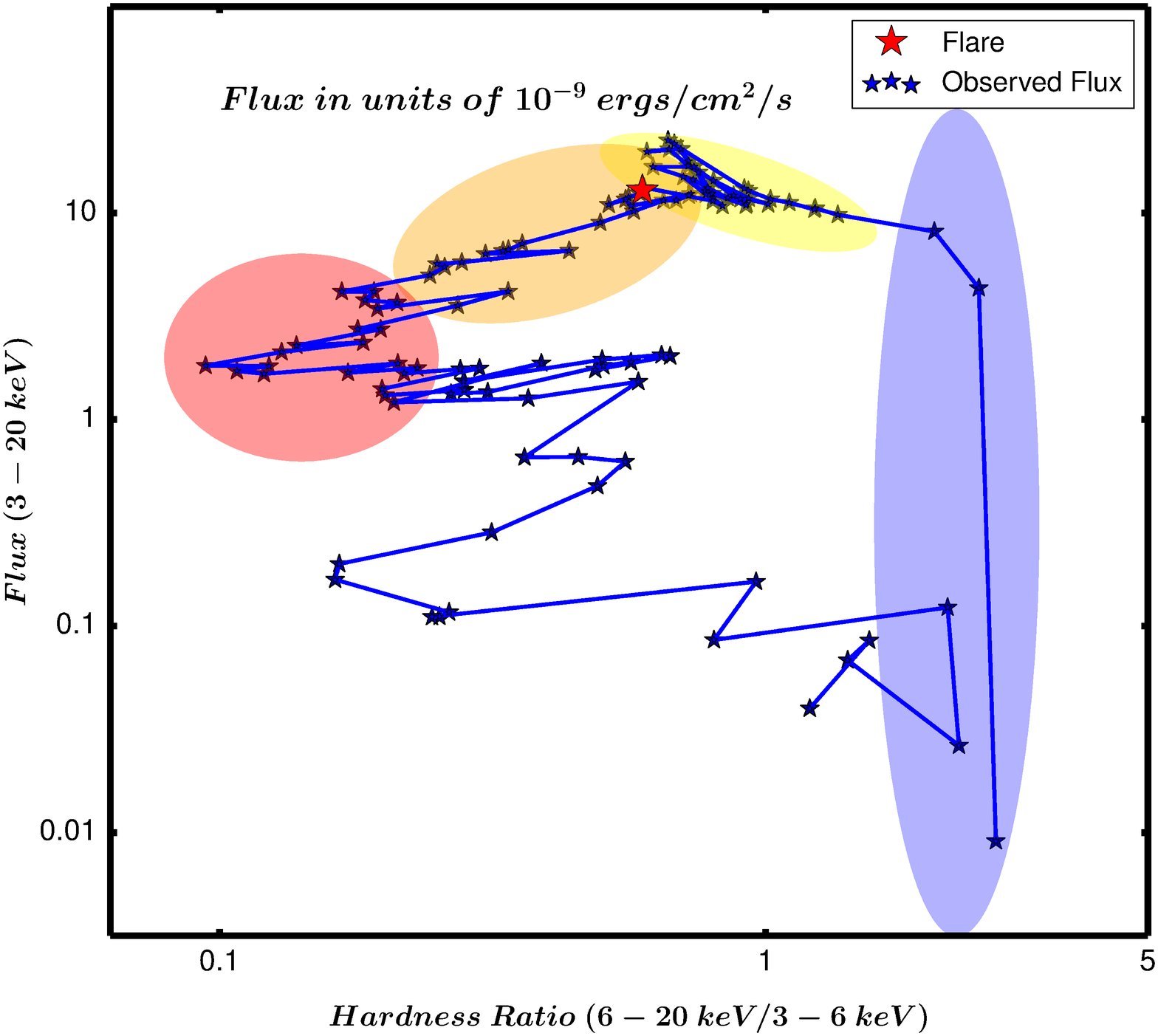} 
\end{flushright}
\caption{The top panel shows the outburst profile of the source XTE J1859+226 during its 1999 
outburst. It is a typical FRED profile with the rising phase extending only around 30 days 
while the decay phase is gradual and takes more than 120 days. The red marker corresponds to 
a radio flare observed during this outburst. The corresponding q-plot is shown in the bottom 
panel. The blue stars indicate the observed flux versus Hardness ratio. The rising phase states 
are marked in coloured patches. The LHS in blue patch, HIMS in yellow patch, SIMS in orange 
patch and HSS in red patch. We also show the occurrence of a radio flare in both outburst 
profile and in HID, marked with a red color star.}
\label{fig:xtej}
\end{figure}

\subsection{Outburst Profile}

Outbursting BH sources either have a Fast Rise Exponential Decay (FRED) or a Slow Rise 
Exponential Decay (SRED) profile. In Figure \ref{fig:xtej}, we present a typical FRED outburst 
profile and the corresponding q-diagram exhibited by the microquasar XTE J1859+226 in the year 
1999. The q-diagram or the HID is the plot of total flux versus hardness ratio (ratio of flux 
in higher energy range to that in lower energy range). BH binaries (BHB) undergo state 
transitions from hard to soft state as the outburst progresses. Hard states correspond to 
observations with considerable amount of high energy photons while the soft states are mostly
dominated by thermal emission with less contribution from high energy photons. A typical 
outbursting BH binary will go through different spectral states (see for details 
Homan {\em et al.} 2001; Belloni {\em et al.} 2005; Remillard \& McClintock, 2006; 
Nandi {\em et al.} 2012) during the outburst phase. The outburst profile and Q diagram for 
XTE J1859+226 is shown in Figure \ref{fig:xtej}. We chose the outburst of XTE J1859+226 as it 
is a `good' representative for the outburst and q-profiles for BHBs. It evolves through all 
the canonical states and hence we have included only this profile in the paper. Similar plots 
for GX 339-4 and IGR J17091-3624 will be presented elsewhere.

As the outburst begins, the source is in LHS where the hardness ratio is maximum. This is followed 
by the HIMS where the photon counts are high and there is significant contribution from the 
higher energies. The source then enters the SIMS where the hardness ratio reduces and the 
soft flux increases. When the source is in the HSS, the hardness ratio reaches its smallest 
value and the spectrum is dominated by thermal emission. In the declining phase, the source 
again occupies the SIMS, HIMS and LHS with similar values for the hardness ratio as was the 
case for the rising phase. In Figure \ref{fig:xtej}, we show the rising phase LHS in blue patch, 
HIMS in yellow patch, SIMS in orange patch and HSS in red patch. After the HSS, the source 
enters the decay phase. The decay phase characteristics are similar to that of the 
corresponding states in the rising phase except for a decrease in total flux values. We also 
note that during the LHS and HIMS of both rising and decay phases, LFQPOs are detected. 
They generally evolve from around 0.1 Hz to a maximum of around 20 to 30 Hz. Here, we study 
the evolution of LFQPO in the rising phase.


\begin{figure*}[hbtp]
\includegraphics[width=18cm,height=6cm]{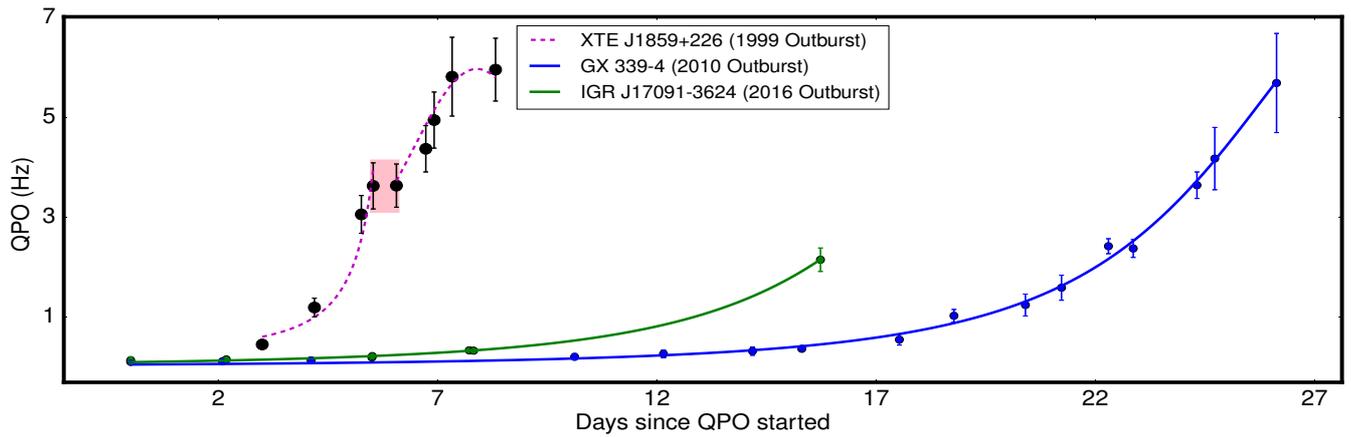} 
\caption{Evolution of QPO frequencies of XTE J1859+226 during its 1999 outburst, GX 339-4 during 
its 2010 outburst and IGR J17091-3624 during its 2016 outburst are shown here alongwith the
fitted model. The shaded region (for XTE J1859+226) shows a QPO frequency which did not evolve 
for around half a day.}
\label{fig:qpoevol}
\end{figure*}

\subsection{Evolution of LFQPOs}

Outbursting BH sources exhibit QPOs which are features that peak in the power spectrum generated 
from source light curves. Typically LFQPOs are of three types C, B and A 
(Casella {\em et al.} 2004). C type LFQPOs appear in the LHS and HIMS only. A and B types are 
of lesser rms values and are seen in the SIMS. C-type QPOs appear in power spectra with flat 
top while the A and B QPOs are found where the power spectra exhibits weak power-law noise. 
In this paper, we study the C-type QPOs, as unlike A and B QPOs they increase in frequency 
during the rising phase of an outburst and decrease in frequency during the decay phase of 
the outburst.

As mentioned above C-type LFQPOs show a time evolution and can be considered to have origins 
based on shock propagation. The Propagating oscillatory shock (POS) model 
(Chakrabarti {\em et al.} 2005; Chakrabarti {\em et al.} 2008; Iyer {\em et al.} 2015) explains 
the time evolution of C-type QPOs as a function of shock location. The shock location is 
essentially the size of the post-shock corona region around the BH. The model considers that 
QPOs are generated by the oscillation of this region. POS model is given by the equation,
\begin{eqnarray}
\nu _{qpo}&=&\frac{c}{2\pi~R~r_g~r_s~\sqrt{r_s -1}}\\
and~~~ ~r_s &=& r_{so} - \frac{v t + \frac{a t^2}{2}}{r_g}\nonumber, \\
\end{eqnarray}
where $\nu _{qpo}$ is the QPO frequency, $R$ is the shock compression ratio or shock strength, 
$v$ is the velocity of the shock front, $a$ is the acceleration, $r_s$ is the instantaneous 
shock location and $r_{so}$ is the initial shock location given in units of $r_g=\frac{2GM}{c^2}$.  

\begin{figure}[!hbtp]
\includegraphics[height=5.75cm,width=0.5\textwidth]{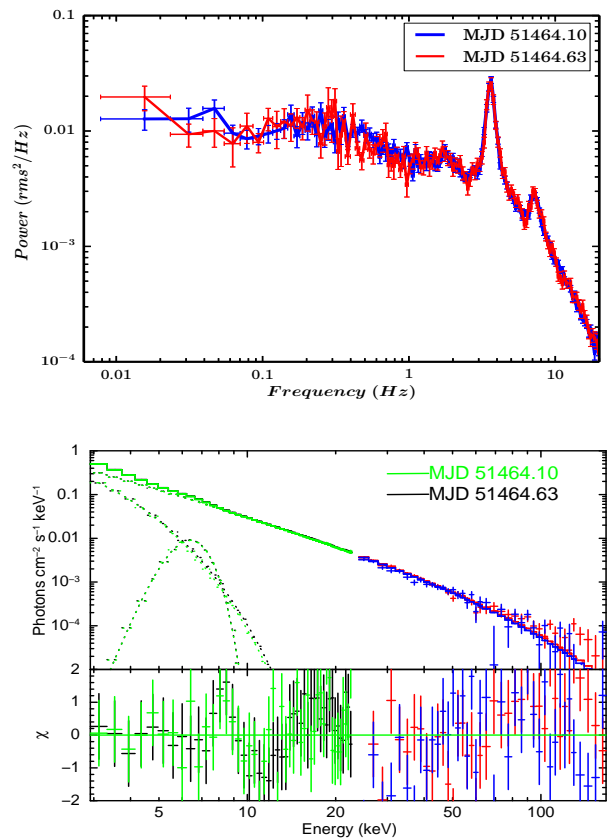} 
\includegraphics[height=8.0cm, width=5.3cm, angle=-90]{comb_enspec.eps} 
\caption{The power spectra (top) and the energy spectra (bottom) of the observations 
corresponding to the shaded region in Figure \ref{fig:qpoevol} are plotted. Here, the green 
and blue lines respectively indicates the PCA and HEXTE spectrum on MJD 51464.10 while the black 
and red lines corresponds to the observation on MJD 51464.63. It is evident that the system 
did not evolve during this period as both energy spectra and power spectra are almost identical.}
\label{fig:combo}
\end{figure}

This model implicitly depends on mass of the compact object and hence can be used to estimate 
the black hole mass from the time evolution of QPOs. Typically the time evolution of QPOs are 
smooth without any discontinuity. In Figure \ref{fig:qpoevol}, we show the case for XTE J1859+226 
during its 1999 outburst. Here, we see that initially the QPO frequency increases with time, 
and then breaks and halts for a while. It then proceeds again with a smooth variation. The two 
frequencies in the shaded region of Figure \ref{fig:qpoevol} corresponding to the days 
MJD 51464.10 and MJD 51464.63 are identical as well as their power spectra are overlapping as 
is shown in Figure \ref{fig:combo}. From here we infer that the QPO parameters like its 
frequency, width, normalization and rms are also identical during this period. Examining the 
energy spectra corresponding to the two observations shaded in pink, we find that the spectral 
parameters are also identical in this period. This suggests that the accretion process has 
been steady and has not evolved from this state for a while. Besides this the QPO evolution 
for this outburst started off with an offset of 3 days as was required for the curve fitting 
using POS model. This indicates that the QPO evolution has started off earlier, but we do not 
have the data as the source was not observed in that period. The other two C-type LFQPO 
evolutions in the Figure \ref{fig:qpoevol} are corresponding to GX 339-4 (2010 outburst) 
and IGR J17091-3624 (2016 outburst). This shows that QPO evolution for different outbursts 
extends for different durations. The QPO evolution of XTE J1859+226 is steep while that for 
GX 339-4 is gradual. Also we notice that in the case of IGR J17091-3624 the maximum frequency 
reached only up till 2.1 Hz while for the other two sources the maximum frequency reached 
around 6 Hz. 

The results of POS modelling for the three sources are presented in Table \ref{tab:posfit}. 
For XTE J1859+226, modelling its QPO evolution gave a shock location of $206~r_g$, initial 
acceleration of $7.42\times 10^{-06}~m/s^2$ followed by a deceleration of 
$1.60\times 10^{-05}~m/s^2$ after the halt and a mass of $5.35\pm0.9~M_{\odot}$. 
Here the modelling is different from Radhika \& Nandi (2014) where they considered two POS 
fits for the two sections of QPO evolution before and after the halt and no acceleration was used.
For the source GX 339-4, 2010 outburst the shock velocity is $10.73 \pm 2~m/s $, constant 
deceleration is $1.45\times 10^{-06}~m/s^2$, the initial shock location is $437~ r_g$ and the 
estimated mass is $10.65\pm 1.67~M_{\odot}$. For the source IGR J17091-3624, we obtained an 
initial shock location of $297 ~r_g$, initial velocity of $9.93 \pm 1.39~m/s$, a constant 
deceleration of $1.81\times 10^{-06}~m/s^2$ and mass of $11.08 \pm 0.79~M_{\odot}$. 
Below, we discuss about the evolution of spectral parameters of different states associated
with outburst phases of the black hole binary sources.

\begin{table*}
    \centering
    \caption{Fit parameters from the POS model}
    \label{tab:posfit}
    \small
    \begin{tabular}{lccccr} %
        \hline
        Source & Outburst & Mass ($M_{\odot}$)&$x_s~(r_g)$ & Initial Velocity $(m/s)$ & Acceleration $(m/s^2)$\\
        \hline
        XTE J1859+226 & 1999 &$5.35\pm 0.9$&$206$&$0 \pm 0.10$&$7.42\times 10^{-06}$,$-1.60\times 10^{-05}$\\
        GX 339-4 & 2010 &$10.65\pm 1.67$&$437$&$10.73 \pm 2$&$-1.45\times 10^{-06}$\\
        IGR J17091-3624 & 2016 &$11.08 \pm 0.79$&$297$&$9.93 \pm 1.39$&$-1.80\times 10^{-06}$\\
        \hline
	\end{tabular}
\end{table*}

\subsection{Evolution of Spectral States}

\begin{figure}[!hbtp]
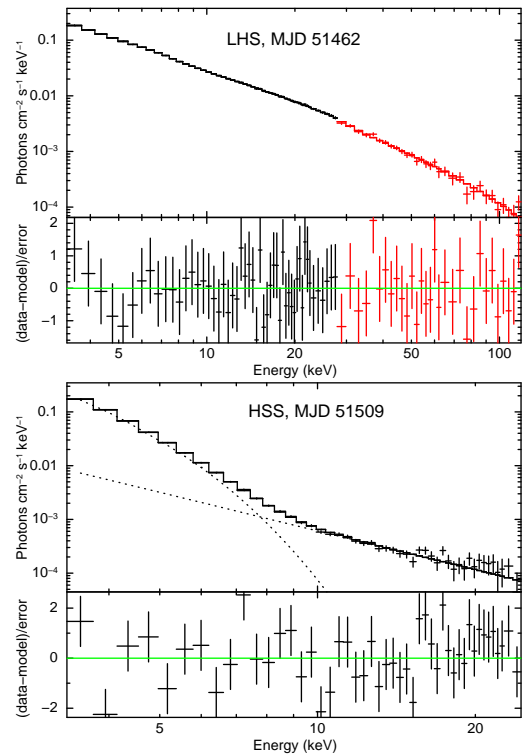

\begin{center}
\includegraphics[scale=0.28,angle=-90]{xtej_pheno_lhs.eps} 
\includegraphics[scale=0.28,angle=-90]{xtej_hss_pheno.eps} 
\end{center}
\caption{Phenomenological modelling of RXTE (PCA-HEXTE) spectra in LHS and HSS of the source 
XTE J1859+226 during its 1999 outburst. No signature of hard X-ray emission in HSS. The black 
lines are spectral data from RXTE-PCA and the red lines from RXTE-HEXTE instrument.}
\label{pheno_xtej}
\end{figure}

Outbursting sources transit through different states as was introduced earlier. The spectral 
evolution of the source XTE J1859+226 has been observed to be similar to typical GBH binaries. 
The spectra could be modelled using disk and \textit{powerlaw} components. The source occupied 
all the spectral states of LHS, HIMS, SIMS and a short duration of HSS, and did complete the 
q-profile in its HID (See Figure \ref{fig:xtej}). The temporal properties revealed the presence 
of type A, B, C and C* (in decay phase) QPOs. Detailed study on this source have been discussed 
in Radhika \& Nandi (2014). 

\begin{figure}[!hbtp]
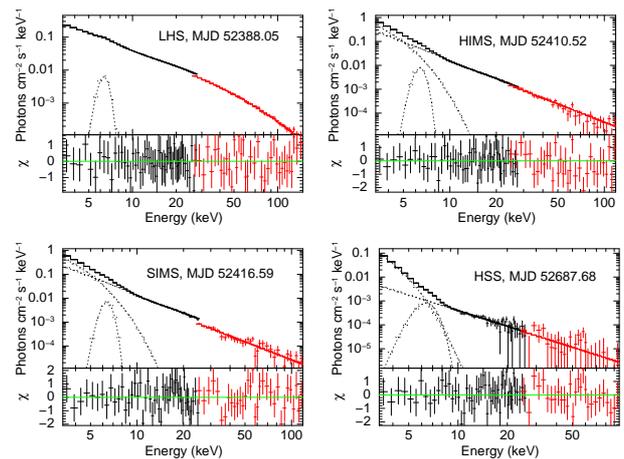

\begin{center}
\includegraphics[scale=0.16,angle=-90]{pheno_lhs_gx2002_modi.eps} 
\includegraphics[scale=0.16,angle=-90]{pheno_hims_gx2002_modi.eps} 
\includegraphics[scale=0.16,angle=-90]{pheno_sims_gx2002_modi.eps} 
\includegraphics[scale=0.16,angle=-90]{pheno_hss_gx2002_modi.eps} 
\end{center}
\caption{Phenomenological modelling of energy spectra corresponding to the four states of LHS, 
HIMS, SIMS and HSS of GX 339-4 during its 2002 outburst. The black curves are from RXTE$-$PCA 
(3$-$25 keV) and the red curves are from RXTE$-$HEXTE (25$-$150 keV). The dotted lines 
indicate the contribution from model components like $diskbb$, $gaussian$ and $powerlaw$.}
\label{pheno_2002_gx}
\end{figure}

The energy spectra from the LHS and HSS of XTE J1859+226 has been shown in 
Figure \ref{pheno_xtej}. The PCA (RXTE) energy spectral data is plotted in black while the 
HEXTE (RXTE) spectral data is plotted in red. The dotted lines indicate the contribution 
from components used in the model like $diskbb$ and $powerlaw$. For the LHS, we fitted the 
spectra with the model \textit{phabs(smedge*cutoffpl)}. The smeared edge was at 7.16 keV and 
the high energy cut-off was at 50 keV. The photon index of the spectrum was $1.61 \pm 0.04$. 
For the HSS, we required a $diskbb$ at 0.75 keV and a $powerlaw$ with index $2.39 \pm 0.18$. 
The HSS spectrum extended only up till 25 keV and no cut-off was required.

\begin{figure}
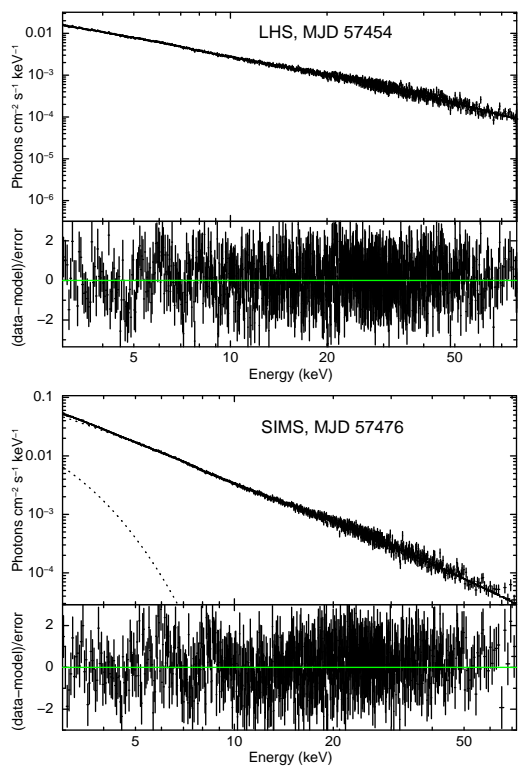

\begin{center}
\includegraphics[scale=0.28,angle=-90]{IGR_march7_pheno.eps} 
\includegraphics[scale=0.28,angle=-90]{IGR_march29_pheno.eps} 
\end{center}
\caption{Phenomenological modelling of NuSTAR spectra in LHS and SIMS of the source 
IGR J17091-3624 during its 2016 outburst. The general model used for fitting the energy spectra
is \textit{phabs(diskbb+ireflect*cutoffpl).}}
\label{pheno_igr}
\end{figure}

We have modelled the energy spectrum of the source GX 339-4 as it went through the LHS, HIMS, 
SIMS and HSS in its 2002 outburst. It was observed that the spectra could be modelled with 
a \textit{powerlaw} alone for the hard states while we required an additional disk 
component (\textit{diskbb}) to model the hard intermediate and soft states. The power-law index, 
$\Gamma$ for the LHS was 1.58 which corresponds to a flat spectrum. For the HIMS and SIMS, 
the photon index was 2.51 and 2.52 respectively and for HSS, $\Gamma$ $\sim$ 2.18. Though the 
value of $\Gamma$ is lower than expected for HSS, it should be noted that the percentage 
contribution of the total flux by \textit{diskbb} in this state was 76.47 \%. The presence 
of \textit{diskbb} in the model indicates a significant contribution of the multi-temperature 
blackbody emission with an inner disk temperature of 0.60 keV and confirms that the source 
is in HSS. Figure \ref{pheno_2002_gx} shows the phenomenological modelling of all four states 
of the source GX 339-4 during its 2002 outburst observed by \textit{RXTE}. 

IGR J17091-3624 during its 2016 outburst was found to occupy the LHS, HIMS and SIMS and only a 
very short presence of HSS (lasting for only a day)(Radhika {\em et al.} 2018). We have included 
two energy spectra from the 2016 outburst of this source corresponding to the LHS and SIMS 
in Figure \ref{pheno_igr}. These are co-ordinated NuSTAR observations. The energy spectra of 
IGR J17091-3624 was modelled using disk, \textit{ireflect} and \textit{cutoffpl} components. 
We obtained a relative reflection value of $0.39 \pm 0.04$ and $0.60 \pm 0.05$ for LHS and SIMS 
respectively. The photon index for LHS was $1.58 \pm 0.01$ and for the SIMS it was 
$2.41 \pm 0.02$. The modelling of simultaneous XRT-NuSTAR observations and details of the entire 
outburst (2016) including state classification, variability and mass estimation will appear 
in Radhika {\em et al.} (2018). 

\subsection{Connection between QPO, Spectral States and Jet Ejection}

Radio jets are a very common phenomena in the black hole accretion process. During `soft' X-ray 
states, the radio emission is strongly suppressed while jets are observed during the `hard' 
states (Fender {\em et al.} 2004; Fender {\em et al.} 2009). Relativistic jets are observed 
as the system changes states from HIMS to SIMS. As a typical example, we have shown the 
case of the source XTE J1859+226 during its 1999 outburst in the Figure \ref{fig:xtej} with the 
location of one of its radio flares marked in red. A detailed study on the radio flares and
connection with spectral states is presented in Radhika \& Nandi (2014). 

We observed that QPOs are absent during the period of multiple jet ejections or flaring of the 
source XTE J1859+226 (Radhika \& Nandi 2014; Radhika {\em et al.} 2016a) have shown that 
type C LFQPOs appear around a day before the jet ejection. There is no presence of QPOs when 
the flare occurs, while type B QPOs are observed a few hours later. During the flares there 
is no phase lag observed between the soft and hard photons. When comparing the flux contributions, 
it has been observed that disk flux dominates over the non-thermal flux during the flares. 
In the LHS and HIMS, the type C QPO frequency increases as both the disk and powerlaw flux 
increases. In the rising phase as well as in the declining phase of the outburst strong radio 
jets have been observed. During the SIMS, it is found that the type B QPO frequency is 
correlated with powerlaw flux, but not with disk flux. This indicates that type B QPOs 
originate from the corona. It has also been noted that a few of the type C QPOs in SIMS are 
not correlated with the powerlaw flux while all C* QPOs are correlated only with the disk 
flux. Besides this the power density spectra had very low rms during the flare. 

Similarly in GX 339-4 during the 2002 and 2010 outbursts, absence of QPOs have been found 
during the time of a radio flare was detected (Radhika {\em et al.} 2016a). The correlation 
of radio flux with X-Ray flux for the source GX 339-4 has been studied in detail by 
Corbel {\em et al.} 2013. Due to lack of radio observations of IGR J17091-3624 during its 
outbursts, we could not explore this characteristic of disk-jet coupling.
 
\subsection{X-Ray Variability}
\begin{figure}[!hbtp]
\hspace{-1cm}
\includegraphics[width=1.0\columnwidth, height=15cm]{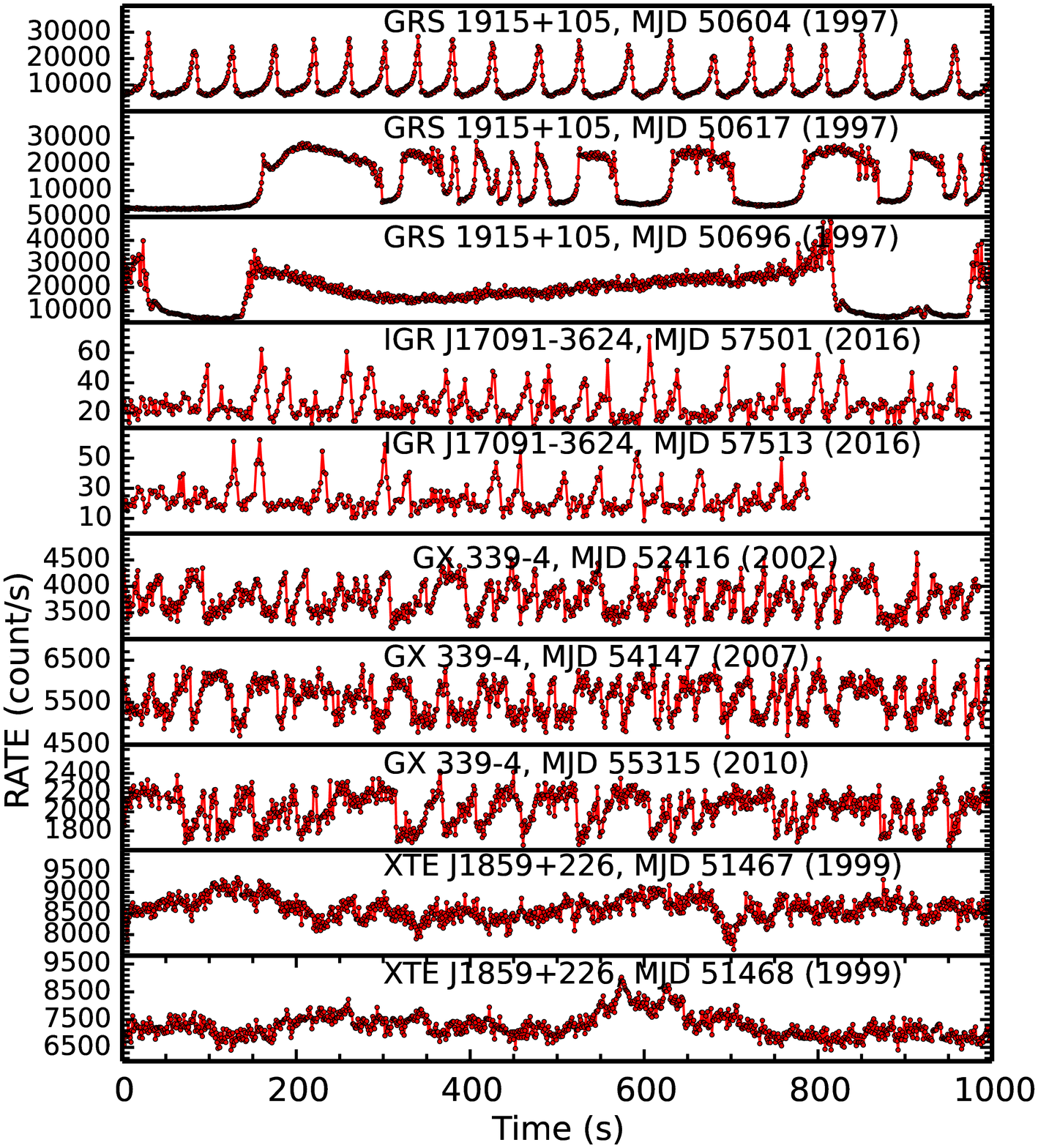} 
\caption{Structured variability exhibited by different BH sources during their outbursts. The 
first three panels show three of the variabilities exhibited by GRS 1915+105 corresponding 
to $\rho$, $\kappa$ and $\theta$ classes respectively. The following panels compare it with 
the variabilities of IGR J17091-3624 (2016), GX 339-4 (2002, 2007 and 2010) and XTE J1859+226 
(1999). The variabilities exhibited by the latter three sources were during the period of 
SIMS which gives us reasonable evidence that GRS 1915+105 probably `locked' in one of its 
intermediate states.}
\label{fig:var_mul} 
\end{figure} 

Next we look into the X-ray variability exhibited by the three sources that we are studying. 
Here, we additionally studied the variabilities of GRS 1915+105 as a reference source as it 
is known for its `class' variabilities. Figure \ref{fig:var_mul} shows the variability 
exhibited by the sources GRS 1915+105 (in 1997), IGR J17091-3624 (in 2016), XTE J1859+226 
(in 1999) and GX 339-4 (in 2002, 2007 and 2010). Belloni {\em et al.} (2001) presented
detailed analysis of the different classes of variability displayed by the source 
GRS 1915+105. We have shown in Figure \ref{fig:var_mul} the $\rho, ~\kappa$ and $~\theta$ class 
variability of GRS 1915+105 respectively. Thus, we see variability at different time scales 
($\sim$ 50 s, $\sim$ 100 s and $\sim$ 600 s) for GRS 1915+105. We compare it with the 
lightcurves of other sources and find that XTE J1859+226 does not have a well defined 
structure in its lightcurves while IGR J17091-3624 and GX 339-4 show some structure or 
repetitions in their lightcurves. The variabilities observed in GX 339-4 are not similar to 
any of the observed `class' in GRS 1915+105. It may be a combination of the 
$\rho$ and $\kappa$ classes exhibited by GRS 1915+105. This requires further investigation.

We did not find any signature of variabilities during the LHS and HIMS. During the SIMS,
we observed that the IGR J17091-3624 exhibits variabilities in the X-ray lightcurve 
(Radhika {\em et al.} 2018) similar to that displayed by the same source in its 2011 outburst 
(Capitanio {\em et al.} 2012; Iyer {\em et al.} 2015). In HSS also the system showed 
oscillations, though weaker than that during the SIMS. It is noticed that IGR J17091-3624 is 
very faint as compared to the other sources that we study. Similarly in XTE J1859+226 and 
GX 339-4, we find signatures of variabilities with different time-scale when the source 
has transited to the SIMS.  

It is also observed that for these sources the variability is observed only in lightcurves 
corresponding to the intermediate spectral states. Thus, we infer that GRS 1915+105 possibly
`locked' in one of its intermediate states as it shows structured variability over the
last two decades.
 
\section{Broadband Spectral Modelling with Two Component Flow - Mass Estimation}

The two component advective flow (Chakrabarti \& Titarchuk 1995; Chakrabarti \& Mandal 2006; 
Iyer {\em et al.} 2015) model incorporates a Keplerian disk (Shakura \& Sunyaev 1973) as well 
as a sub-keplerian halo around the black hole. The Keplerain disk is in the equatorial region 
and the sub-Keplerian halo is on top and bottom sides of the disk. During the hard states, 
the sub-Keplerian component predominates the flux contribution while during the soft states 
the major flux contribution is from the Keplerian disk. Here unlike in phenomenological models, 
the high and low energy photon emission are dependent on each other. As the outburst progresses, 
the effect of the sub-Keplerian component reduces and the disk starts contributing more to 
the total flux of the source. 

We self-consistently calculate the total radiation spectrum from hydrodynamics 
(Mandal \& Chakrabarti 2005) and import the two component advective flow model as an 
additive table in \textit{XSPEC} (Iyer {\em et al.} 2015). For each source under consideration 
there are four variable parameters in this model. They are the mass of the black hole $(M_{BH})$ 
in units of $M_{\odot}$ , shock location $(r_s)$ in units of $r_g$, keplerian disk accretion 
rate, $\dot{m} _d$ and sub-keplerian halo accretion rate ($\dot{m}_h$) in units of Eddington 
rate. Besides this we have the \textit{norm} parameter which is a constant for a source as 
it is dependent on distance to the source, its mass and inclination.
For the fitting, we use spectra from 3 to 150 keV. During the softer states as the higher 
energy range of the spectral data decreases we fit only up to 80 keV or so. We perform fitting 
with the two component model for all data sets of an outburst of the source leaving the `norm' 
parameter free. Then we take an average of all the norms and fix it as the norm for the source. 
Fit results for both phenomenological and two component models for all three sources under 
consideration are presented in Table \ref{tab:allfit}.

\begin{figure}[!hbtp]
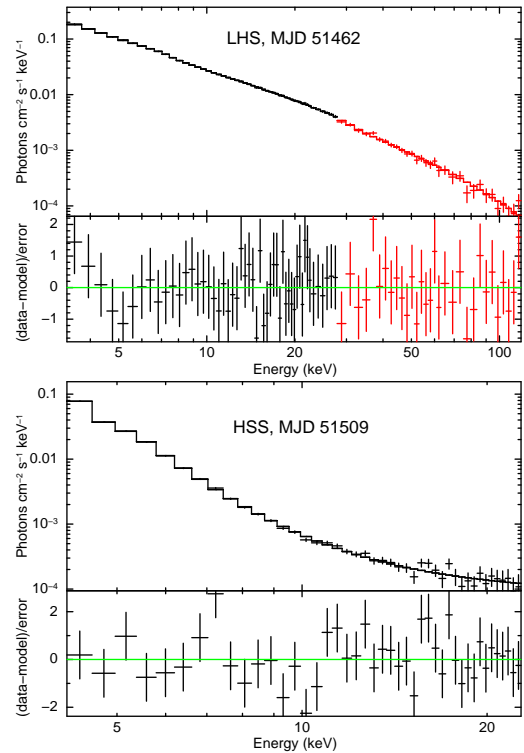

\begin{center}
\includegraphics[scale=0.28,angle=-90]{xtej_tcaf_lhs.eps} 
\includegraphics[scale=0.28,angle=-90]{xtej_hss_tcaf.eps} 
\end{center}
\caption{Two component flow modelling of RXTE (PCA-HEXTE) spectra in LHS and HSS of the source 
XTE J1859+226 during its 1999 outburst.}
\label{tcaf_xtej}
\end{figure}

Figure \ref{tcaf_xtej} shows the energy spectra of the source XTE J1859+226 during its 
1999 outburst modelled with the two component flow. We present results only from two states 
(LHS and HSS) in this paper. Detailed analysis of each state of the outburst is presented in 
Nandi {\em et al.}(2018). When the source was in the LHS (on MJD 51462), it had a halo accretion 
rate of 0.26 $\dot{M}_{Edd}$ and a Keplerian accretion rate of 0.15 $\dot{M}_{Edd}$. The shock 
location was at $101 \pm 2~r_g$ and the mass of the source is estimated as 
$6.0 \pm 0.53~ M_{\odot}$. In the HSS, the accretion rates are 0.18 $\dot{M}_{Edd}$ for the halo 
and 0.49 $\dot{M}_{Edd}$ for the disk. The computation of shock location gave a value of 
$24.6 \pm 1.2~r_g$. The mass estimate for the source from this model is 
$6.05 \pm 0.27~ M_{\odot}$. 

\begin{figure}[!hbtp]
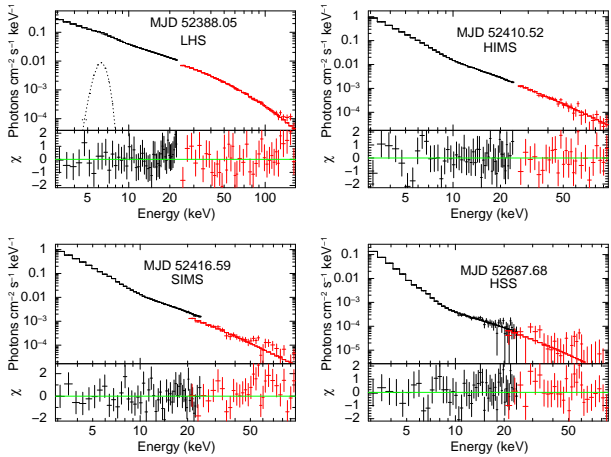

\begin{center}
\includegraphics[scale=0.16,angle=-90]{2002_GX339_LHS_TCAF_modi.eps}  
\includegraphics[scale=0.16,angle=-90]{2002_GX339_HIMS_TCAF_modi.eps} 
\includegraphics[scale=0.16,angle=-90]{2002_GX339_SIMS_TCAF_modi.eps}
\includegraphics[scale=0.16,angle=-90]{2002_GX339_HSS_TCAF_modi.eps}  
\end{center}
\caption{Two component flow modelling of energy spectra corresponding to the four states of 
GX 339-4 during its 2002 outburst. The black lines correspond to spectral data from PCA, the 
red lines correspond to spectral data from HEXTE and the dotted lines are the model components 
like \textit{gaussian} and two component flow.}
\label{tcaf_2002_gx}
\end{figure}

\begin{figure}[!hbtp]
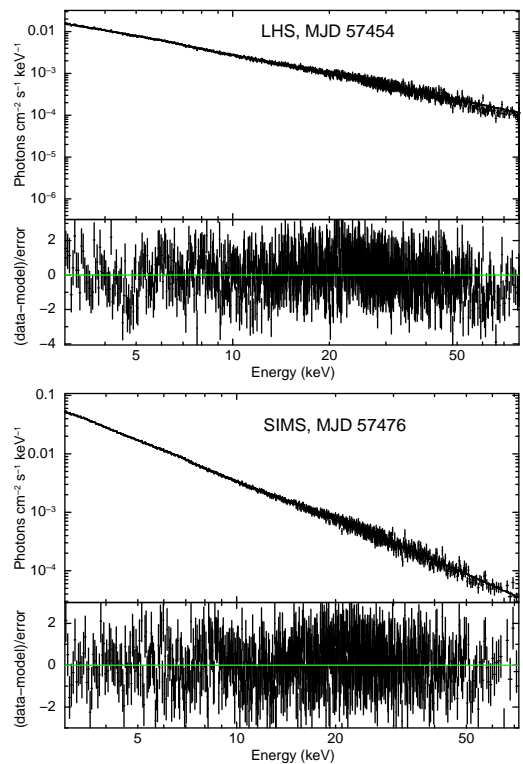

\begin{center}
\includegraphics[scale=0.28,angle=-90]{IGR_march7_tcaf.eps} 
\includegraphics[scale=0.28,angle=-90]{IGR_march29_tcaf.eps} 
\end{center}
\caption{Two component flow modelling of NuSTAR spectra in LHS and SIMS of the source 
IGR J17091-3624 during its 2016 outburst.}
\label{tcaf_igr}
\end{figure}

\begin{table*}
    \centering
    \caption{Spectral fit parameters using phenomenological and two component flow models}
    \label{tab:allfit}
    \tiny
    \begin{tabular}{lcc|cccc|ccr} %
        \hline
        Source & MJD (State) & Observatory&Tin (keV) & $rel\_refl$ & Photon Index&high-cut (keV)&$\dot{m_h},\dot{m_d}~(\dot{M}_{Edd})$&Mass ($M_{\odot}$)&$x_s~(r_g)$\\
        \hline
        &&&& Phenomenological & & && Two component flow &\\
        \hline
        XTE J1859+226 & 51462 (LHS) & PCA-HEXTE & - & - & $1.61 \pm 0.04$&$50 \pm 8.0$&$0.26,~0.15$&$6.0\pm 0.053$&$101\pm 2$\\
        XTE J1859+226 & 51509 (HSS) & PCA&$0.75 \pm 0.005$& - & $2.39 \pm 0.18$&-&$0.18,~0.49$&$6.05 \pm 0.27$&$24.6\pm 1.2$\\
        GX 339-4 & 52388 (LHS) & PCA-HEXTE & - & $0.39\pm 0.17$ & $1.58 \pm 0.04$ &$71\pm 6$&$0.20,0.04$&$11.56 \pm 0.17$ &$424 \pm 7$\\
        GX 339-4 & 52410 (HIMS) & PCA-HEXTE &$0.86 \pm 0.01$ & - & $2.52\pm 0.02 $& - &$0.06,3.3$&$11.02 \pm 0.14$&$190 \pm 12$\\
        GX 339-4 & 52416 (SIMS) & PCA-HEXTE &$0.88 \pm 0.03$ & - & $2.51 \pm 0.07$ & - &$0.04,0.57$&$10.21 \pm 0.35$&$46 \pm 2$\\
        GX 339-4 & 52687 (HSS) & PCA-HEXTE & $0.60 \pm 0.05$ & - & $2.18 \pm 0.11$ & - &$0.64,10.6$&$9.55 \pm 0.47$&$5 \pm 0.03$\\
        IGR J17091-3624&57454 (LHS)&NuSTAR&$-$ &$0.39^{+0.06}_{-0.06}$&$1.58^{+0.028}_{-0.028}$&$228^{+72}_{-44}$&$0.36 ,~0.04 $&$11.66 \pm 0.12$&$484 \pm 18$\\
        IGR J17091-3624&57476 (SIMS)&NuSTAR&$0.59^{+0.014}_{-0.015}$ &$0.60^{+0.07}_{-0.08}$&$2.41^{+0.017}_{-0.020}$&$-$&$0.07 ,~0.19 $&$11.51 \pm 0.07$&$56 \pm 2$\\
        \hline
	\end{tabular}
\end{table*}

In Figure \ref{tcaf_2002_gx}, we have shown the two component flow based spectral modelling 
performed for GX 339-4 during its 2002 outburst. Here, we find that during the LHS the 
shock location is 424 $\pm$ 7 $r_g$, halo rate is 0.20 $\dot{M}_{Edd}$ and disk rate is 
0.04 $\dot{M}_{Edd}$. As the source moves to HIMS, the shock location becomes 190 $\pm$ 12 $r_g$. 
The halo rate declines to 0.06 $\dot{M}_{Edd}$ and while disk rate increases to 
3.3 $\dot{M}_{Edd}$. In SIMS, the shock location has reduced further to 46 $\pm$ 2 $r_g$, the 
halo rate is 0.04 $\dot{M}_{Edd}$ and the disk rate increased to 0.57 $\dot{M}_{Edd}$. Finally, 
in the HSS we obtained the limiting value of shock location of 5 $r_g$, a halo rate of 0.64 and a 
disk rate of 10.6 both in units of $\dot{M}_{Edd}$. We also estimated the mass from spectral 
modeling with two component flow and found that GX 339-4 mass lies in the range 9.20 to 
11.95 $M_{\odot}$.
Heida {\em et al.} (2017) has estimated the black hole mass of GX 339-4 to be between 
$2.3 ~M_{\odot}$ and $9.5 ~M_{\odot}$. Detailed broadband spectral modelling of all outbursts 
of GX 339-4 is under progress and the results will be presented elsewhere.

Similarly, we have studied the source IGR J17091-3624 during its 2016 outburst by means of 
two component flow modelling of SWIFT and NuSTAR simultaneous spectra (Radhika {\em et al.} 2018). 
During the LHS, the shock location obtained was $484~r_g$ and it reduced to $56~r_g$ as the 
source entered the SIMS. The halo accretion rate changed from 0.36 to 0.07 $\dot{M}_{Edd}$ and 
the disk accretion rate increased from 0.04 to 0.19 $\dot{M}_{Edd}$. The mass estimates from 
the two observations are $11.66 \pm 0.12$ and $11.51 \pm 0.07 ~M_{\odot}$. 

Figure \ref{tcaf_igr} shows the energy spectra of IGR J17091-3624 during the LHS and SIMS of 
its 2016 outburst as observed by NuSTAR. Thus we could understand the spectral evolution of 
the different sources based on two component advective flow modelling and estimate the mass 
of the compact objects in these systems. 

\section{Conclusion and Future Work}

In this paper, we present the results of evolution of spectro-temporal characteristics of 
three different GBH binaries. First, we present the frequency of outbursts for different sources 
that are quite different from each other. Some sources like GX 339-4 are dynamic whereas sources 
like XTE J1859+226 remains in a quiescent phase for most of the time. The evolution of outburst
profile of these sources show that they do not have a unique variation of the intensity. 
The profiles are found to be either a FRED or slow rise and exponential decay. 

We have studied how the black hole binary sources evolve along the hardness intensity diagram 
(q-plot) during their outburst. The sources XTE J1859+226 and GX 339-4 are observed to complete 
the q-profile and hence exhibit all the spectral states. IGR J17091-3624 is found to be 
different from this typical variation and do not occupy all the spectral states. The 2016 
outburst of IGR J17091-3624 has just one observation corresponding to a possible HSS wherein 
the hardness ratio was a minimum and the signature of disk was prominent 
(Radhika {\em et al.} 2018). 

We have also modelled the energy spectra using the two component flow model and found the 
variation of spectral characteristics to be matching with that obtained using 
phenomenological modelling. With the help of this, we have been able to find that the 
shock location decreases as a source evolves from its LHS to HSS. Additionally, the variation 
of halo rate and disk rate agrees well with the variation of soft and hard fluxes respectively. 
During the LHS, the halo rate is found to be maximum and decreases as the source approaches 
HSS. Similarly, the disk rate increases when the source energy spectra softens towards the 
HSS (Radhika {\em et al.} 2018). 

The temporal characteristics show the presence of LFQPOs in the power spectra. Different types 
like A, B, C and C* have been identified for the different sources. The C type QPO frequency is 
observed to increase as the outburst progresses. We also modelled the time evolution with the 
help of the POS model and understand that the QPO is generated by the propagation of a shock 
front (see Figure \ref{fig:qpoevol}).  

There also exists a possible connection between the occurrence of radio flares and 
spectro-temporal characteristics of the sources. 
Figure \ref{fig:xtej} shows the location of radio flare for the 1999 outburst of the source 
XTE J1859+226. We find that during the period the flare occurs as indicated by radio lightcurves, 
the power spectra does not show presence of QPOs and the total rms variability decreases 
(Radhika \& Nandi 2014). The energy spectra is observed to get softer as evident in increase 
of soft flux (disk rate) and decline in the hardness ratio. This suggest that probably the 
innermost part of the disk (hot corona) is being evacuated into jets/flares, resulting in the 
soft flux to dominate. Hence, the absence of any oscillations of the corona leads to no 
detection of QPOs. Thus, we can be certain that the origin of QPOs is linked with the 
oscillations of the corona.

An interesting feature observed in all the sources studied in this paper, is the occurrence 
of variability signatures in the X-ray lightcurves. These are found to be similar to that 
observed in GRS 1915+105 (Belloni {\em et al.} 2001) and IGR J17091-3624 (2011 outburst; 
Altamirano {\em et al.} 2011). We find that the variability signatures are not identical for 
all the sources studied here. Additionally, a comparative study with the spectral evolution 
suggest that these variabilities appear at the time the source transits to the intermediate 
state (specifically SIMS in XTE J1859+226 and GX 339-4). We consider this as an indication 
that GRS 1915+105 which exhibits different `class' variabilities is probably `locked' in one 
of the intermediate states.

Finally, we also modelled the energy spectra of the three sources using the two component 
advective flow paradigm and estimated the Keplerian and sub-Keplerian accretion rates, 
shock location and mass of the compact objects. These results are presented in 
Table \ref{tab:allfit}.

In this paper, we have not considered the effects of black hole spin in order to understand 
the spectro-temporal evolution. As a future study we intend to do the analysis of QPOs and 
jet ejection mechanisms in the Kerr space. Broadband analysis and modelling of the energy 
spectra considering the rotational effects of the compact object can shed new light on the 
origin of QPOs. Moreover, the effect of black hole spin on the radio flaring of the BHBs will 
also be studied in detail.

\section*{Acknowledgements}

We are thankful to the reviewer whose valuable suggestions have helped in improving the 
manuscript. AN thanks GD, SAG; DD, PDMSA and Director, ISAC for encouragement and continuous 
support to carry out this research. This research has made use of the data obtained through 
High Energy Astrophysics Science Archive Research Center online service, provided by 
NASA/Goddard Space Flight Center.

\end{document}